# Current-tunable room temperature ferromagnetism and current-driven phase transitions

Jianping Guo[1], Peng Rao[1], Xinhao Huang[1], Tailai Xu[1], Yuxuan Guo[1], Jian Shao[1], Cheng Sun[2,3], Anton Orekhov[2], Thomas N. G. Meier[1], Johannes Knolle[1], Christian H. Back[1], and Lin Chen[1]

[1]*Department of Physics, Technical University of Munich, Munich, Germany*

[2]*Department of Chemistry, Technical University of Munich, Munich, Germany*

[3]*TUMint.Energy Research GmbH, Department of Chemistry, Technical University of Munich, Munich, Germany.*

**It is generally assumed that the application of a charge-current in ferromagnetic metals suppresses their ferromagnetic order through trivial Joule heating. Here, we demonstrate that a charge current can instead *enhance* magnetic ordering. Using a WTe$_2$/Fe$_3$Ge$_2$Te (FGT) stack as a model system, we show that a charge current flowing in WTe$_2$ controls the ferromagnetic properties and magnetic phase transition of the adjacent FGT via a current-induced effective magnetic-field arising from orbital magnetization. Remarkably, the charge current drives a substantial enhancement of the Curie temperature, boosting it well above room temperature. Furthermore, we show that the charge-current enables controlled tuning of the phase transitions in FGT, which confirms the scaling behaviour of a ferromagnet-paramagnet phase transition. This work provides a pathway for**

**integrating two-dimensional ferromagnets into spintronic functionalities at technologically relevant temperatures and for exploring novel current-driven phenomena in ferromagnetic systems.**

The discovery of magnetic order in two-dimensional van der Waals materials[1-6] provides an ideal platform to study magnetism in reduced dimensions and to develop exotic spintronic functionalities beyond conventional ferromagnetic transition metals. Various important spintronic phenomena have been realized based on van der Waals magnets (vdWMs), including tunnelling magnetoresistance[7], spin-filtering effects[8-10], electric-field control of magnetism[11-14], spin-torque effects[15-17] and spin wave excitations[18-19]. However, most device operations occur at temperatures far below room temperature. For practical applications, it is crucial to explore robust and efficient routes to raise the Curie temperature ($T_C$) of vdWMs above room temperature.

Because the lattice constants of vdWMs are much larger than those of conventional ferromagnetic metals (e.g., Fe, Co and Ni), the exchange interaction in vdWMs is significantly weaker, leading to a much lower $T_C$ in the bulk form. To date, only a few vdWMs exhibit $T_C$ near or above room temperature (Ref. 20 and Supplementary Section 1). Moreover, a pronounced reduction of $T_C$ is commonly observed in the monolayer or few-layer limit[6,12,21] arising from the combined effect of enhanced thermal fluctuation by spin wave excitation, reduced exchange interaction, as well as reduced magnetic anisotropy due to the absence of neighbouring layers along the out-of-plane direction. Due to these shortcomings, it is highly desirable to develop robust and efficient approaches to enhance the $T_C$ of vdWMs, particularly in the monolayer or few-layer regime since most devices are built from thin film heterostructures. Furthermore, since practical devices are operated by either current or voltage, electrical control of magnetism is of central importance[22]. Previous studies have demonstrated that electric fields generated in capacitors or gating structures can significantly enhance magnetism in vdWMs by modulating the exchange interaction, including the magnetic insulators such as $CrI_3$[13,14] and $Cr_2Fe_2Te_6$[11], as well as the metallic ferromagnet $Fe_3GeTe_2$ (FGT)[12]. Note that the gating approach relies on charge transfer, and the efficiency depends on the charging/discharging capacity.

On the other hand, the physics and the universality of phase transition in low-dimensional systems have long been subjects of intense interest[23]. At $T_C$, the isothermal ferromagnet–paramagnet transition is typically driven by an external magnetic field $H$, which serves as the control parameter. The dependence of the order parameter (e.g., magnetization) on $H$ at $T_C$ has been extensively studied both theoretically and experimentally across a wide range of systems and models. However, the external magnetic field couples uniformly to the entire sample. It is therefore important to explore whether a *local* control parameter can be used to drive magnetic phase transitions.

Here, we report a distinct mechanism of electrical-current control of magnetism that does not rely on extra charge transfer. Instead, it exploits the current-induced out-of-plane magnetization $\mathbf{m}_z^I$ in a *bilayer* of $WTe_2$ (1 layer = 0.8 nm), which couples to an adjacent FGT layer via the magnetic proximity effect. Note that here the magnetic proximity effect refers to the current-induced magnetic moment in WTe$_2$ polarizing the paramagnetic FGT, which is different from the conventional definition of the magnetic proximity effect, i.e., a ferromagnet induces a magnetic moment in non-magnetic materials due to interfacial exchange coupling. We demonstrate robust and reversible electrical control of $T_C$, with a charge current of 0.5 mA raising $T_C$ from approximately 200 K in the pristine FGT to 370 K. Furthermore, we show that the charge current serves as a new control parameter to drive the paramagnetic-ferromagnetic phase transition, and that universal scaling behaviour is confirmed experimentally.

We fabricate WTe$_2$/FGT stacks by mechanical exfoliation (Fig. 1a and Supplementary Section 2). Here the thickness of WTe$_2$ is fixed to the bilayer since bilayers of WTe$_2$ host a sizeable Berry curvature dipole (BCD)[24-30]. The BCD gives rise to the generation of $m_z^I$ under a dc current $I$ applied along the low-symmetry $a$-axis of WTe$_2$, i.e, $m_z^I = cI$, where $c$ characterizes the charge-spin conversion efficiency. FGT is a metallic vdWM that retains perpendicular magnetic anisotropy down to the monolayer limit[6,12]. Fig. 1b presents the high-resolution transmission electron microscopy (HRTEM) image of a thicker WTe$_2$/FGT device, revealing a sharp and atomically smooth interface between WTe$_2$ and FGT.

As shown in Fig. 1c, the local magnetization of the thin FGT flake is probed via the anomalous Hall effect (AHE). Empirically, the anomalous Hall resistivity ($\rho_{AH}$) is proportional to the out-of-plane spontaneous magnetization $M_z$[31],

$$\rho_{AH} = R_S\mu_0 M_z, \qquad (1)$$

where $\mu_0$ is the magnetic constant and $R_S$ the anomalous Hall coefficient. Here, an ac current $I_{ac}$ and a control dc current $I$ are applied *simultaneously* along the $a$-axis of WTe$_2$, and the transverse ac voltage $V_{ac}$ is detected by lock-in amplifier. The amplitude and the frequency of $I_{ac}$ are 10 μA and 31 Hz, respectively. Fig. 1d shows Hall resistivity loops $\rho_{xy}(H_z)$ measured by sweeping the out-of-plane magnetic field $H_z$ for $I = 0$ for WTe$_2$ (1.6 nm)/FGT (2.4 nm) from 150 K to 200 K, where $\rho_{xy} = (V_{ac}/I_{ac})t_{FGT}$. Well-defined square hysteresis loops are observed at each temperature, indicating that the perpendicular magnetic anisotropy of FGT is still preserved. The coercive force decreases monotonically with increasing temperature and vanishes at approximately 200 K. Fig. 1e presents the temperature dependence of the anomalous Hall resistivity $\rho_{AH}$, which is extracted from the loop height (height = $2\rho_{AH}$). $T_C$ for this thin FGT is slightly lower than that of thicker ones, which is possibly due to the finite size effect previously observed in ultrathin 3d ferromagnetic metals[32]. Notably, the presence of WTe$_2$ strongly influences the $T$-dependence of $\rho_{AH}$, giving rise to a non-monotonic behavior below 80 K. This behaviour is absent in pure FGT as shown in the inset of Fig. 1e and cannot be explained by the conventional expression for AHE (Equation 1). This indicates that the AHE doesn't directly detect the temperature dependence of the magnetization in the WTe$_2$/FGT stack. Similar behaviour has also been reported in other ferromagnets and is widely attributed to a dominant intrinsic Berry curvature contribution to the AHE[33].

Previous studies on a similar heterostructure have reported signature of the topological Hall effect in the $\rho_{xy}$ loops of WTe$_2$/FGT bilayer, suggesting a possible proximity-induced Dzyaloshinskii–Moriya interaction arising from WTe$_2$[34]. However, after examining more than ten devices over a wide range of temperatures, WTe$_2$ and FGT thicknesses, we observe no evidence of topological Hall contributions (Supplementary Section 3).

Having established the magneto-transport properties of the WTe$_2$/FGT stack below $T_C$, we now turn to the electrical-current modulation of magnetism above $T_C$. As shown in Fig. 2a, an dc current $I$ with varying amplitude is applied at $T = 300$ K. For $I = 0$, the anomalous Hall conductivity $\sigma_{AH}$ remains zero, consistent with FGT being in the paramagnetic phase. Upon applying a positive $I$, a finite positive $\sigma_{AH}$ emerges and remains stable until the current polarity is reversed, upon which $\sigma_{AH}$ also switches sign. The $I$-dependence of $\sigma_{AH}$ is summarized in Fig. 2b. The magnitude of $\sigma_{AH}$ scales linearly with $I$ for $|I| \leq 0.3$ mA, while deviations from linearity at higher currents are attributed to Joule heating. Because $\sigma_{AH}$ is a hallmark of ferromagnetic order, these results provide strong evidence that ferromagnetic order is induced by the charge current at $T = 300$ K.

A recent theoretical study proposed that current-induced spin-orbit torques in ultrathin FGT could reorient the magnetization from out-of-plane to the in-plane direction and potentially enhance its $T_C$[35]. To test this scenario, we performed identical measurements on a single FGT film without WTe$_2$. As shown in Fig. 2c, $\sigma_{AH}$ remains zero for all applied currents, with no evidence of current-generated AHE response. This demonstrates that the observed current-induced magnetism in WTe$_2$/FGT bilayer originates from the presence of WTe$_2$, rather than from FGT alone.

The current induced ferromagnetic order can be understood as follows. For $T > T_C$ (200 K), thermal perturbations destroy long-range ferromagnetic order and lead to the disappearance of $\sigma_{AH}$. However, the charge current in WTe$_2$ generates $m_z^I$ via the Berry curvature dipole and $m_z^I$ can effectively influence the adjacent FGT via the magnetic proximity effect. As schematically illustrated in Figs. 2d-2f, this additional current-induced exchange interaction $\Delta$ overcomes the thermal perturbation and leads to a finite $\sigma_{AH}$ in FGT. Note that both, the direction and the magnitude of $m_z^I$ in WTe$_2$ are determined by the polarity and the magnitude of the applied current, and that the resulting $\Delta$ and $\sigma_{AH}$ are expected to be odd functions with respect to the current. This exactly reflects the odd-in-current symmetry of the AHE response.

Figure 3a shows the $T$-dependence of $\sigma_{AH}$ for $I$ from +0.5 mA to −0.5 mA. For $I = 0$ mA, $\sigma_{AH}$ vanishes at approximately 200 K, consistent with the $T_C$ of pristine FGT. Below 200 K, $\sigma_{AHE}(+I) \neq \sigma_{AHE}(-I)$ holds, reflecting the current-induced modulation of AHE, in agreement with a previous report[36]. Above 200 K, the magnitude of the current-induced $|\sigma_{AHE}|$ decreases quasi-linearly with increasing $T$. To quantify the current-induced $T_C$, we fit the $\sigma_{AHE}(T)$ traces with linear functions (dashed lines) and extract $T_C$ from $T$-axis intercepts (solid symbols). As shown in Fig. 3b, positive and negative currents result in identical transition temperatures, satisfying $T_C(-I) = T_C(+I)$. The extracted $I$-dependence of $T_C$ is summarized in Fig. 3b and shows a monotonic increase with increasing $|I|$. Notably, a current of 0.5 mA raises the transition temperature to approximately 370 K, nearly twice the intrinsic $T_C$ of the pristine FGT flake (~ 200 K). Moreover, current-orientation dependent measurements (Supplementary Section 4) show that when the charge current is applied along the high symmetry $b$-axis, no $T_C$ modulation is observed, providing strong evidence that the Berry curvature dipole plays a dominant role in the current-induced paramagnetic-ferromagnetic phase transition.

We note that for conventional electric-field control in capacitor structures, the modulation of $T_C$ is unidirectional with respect to the gate-voltage $V_G$, i.e., $T_C(-V_G) \neq T_C(+V_G)$ usually holds[12,22]. In contrast, the modulation by electrical current control is bipolar, i.e., $T_C(-I) = T_C(+I)$ holds. Moreover, Figs. 2a and 3c demonstrate that one can transform a paramagnetic state into a ferromagnetic state simply by applying a charge current, with the magnitude of the current-induced magnetization linearly proportional to $I$. Beyond enabling a paramagnetic-ferromagnetic phase transition above room temperature, the direction of the induced ferromagnetic order can be reversed by reversing the polarity of **I**. Therefore, this approach introduces a new mechanism for writing magnetization information beyond the well-known spin-transfer torque induced magnetization switching[37-40]. Although the current-induced magnetization is volatile for $T > 200$ K – disappearing once the current is off – this telegraph-like signal may offer promising opportunities for writing magnetic information. Similar step-like, polarity-dependent responses are observed in ferroelectric switching and memristive systems, where the controlled redistribution of internal states underlies the functionality[41].

The $\sigma_{AH}$-$T$ traces under different modulation currents shown in Fig. 3a resemble $M$-$T$ curves measured under external magnetic field, where a tail is formed by the application of $H$ and the response depends on the magnitude of $H^{23}$. In our case, the charge current replaces $H$ as a control parameter and removes the ferromagnet-paramagnet phase transition. To test this assumption, we quantify the isothermal critical exponent $\delta$ as follows. Because AHE does not directly probe the $T$-dependence of the order parameter $M$, we assume that, in the vicinity of the critical point, $\sigma_{AH} \propto M^\alpha$ holds.

Around $T_C$, the temperature dependence of $M$ follows $M \propto (-t)^\beta$, with $t = \frac{T}{T_C} - 1$, and the exponent $\beta$ (= 0.37) has been determined by magnetization[42,43] and magnetic circular dichroism measurements[6]. Therefore, the temperature dependence of $\sigma_{AH}$ at $I = 0$ is fitted by $\sigma_{AH}(t, I=0) \propto (-t)^{\alpha\beta}$, and from the fitting we obtain $\alpha = 2.1$ and $T_C = 197$ K (Fig. 4a). At the critical temperature, the $I$-dependence of $\sigma_{AH}$ is fitted by $\sigma_{AH}(t=0, I) \propto I^{\frac{\alpha}{\delta}}$, and $\delta$ is determined to be 3.10 (Fig. 4b). This current-induced critical exponent is close to the mean-field value ($\delta = 3$) (Supplementary Section 5) and is significantly smaller than the $H$-induced critical exponent determined by magnetization measurements in bulk samples ($\delta = 4.3$–$4.5$)[42,43]. The much smaller $\delta$ value determined here may arise from the following factors: i) our FGT thickness (2.4 nm) is much thinner than bulk samples, and the dimension of the sample should influence the critical exponent, ii) in our device, the current-induced exchange field only acts on FGT locally at the interface, while the $H$-field homogeneously acts on the whole ferromagnet, iii) Joule heating cannot be avoided in our measurements, and thus the $\sigma_A$-$I$ trace shown in Fig. 4b is not strictly an isothermal transition. Nevertheless, we respectively scale $\sigma_{AH}$ and $t$ by $I^{1/\delta}$ and $I^{1/\beta\delta}$ [44,45]. As plotted in Fig. 4c, all the $\sigma_{AH}(T, I)$ traces collapse onto one single curve, which proves the scaling hypothesis for phase transitions. Based on these results, we conclude that the charge current applied in the WTe$_2$/FGT device can behave as a control parameter for magnetic phase transitions.

Finally, we develop a theory of the temperature dependence of the modulation in order to account for the key qualitative signatures. We consider the two-band model that takes into account the parity anomaly of AHE[31,46]

$$h(k) = \frac{\hbar^2 k^2}{2m} - \mu + v(\mathbf{k} \times \boldsymbol{\sigma}) \cdot \mathbf{z} - (M + \Delta)\sigma_z \tag{2}$$

Here $k$ is the Bloch momentum, $m$ the effective mass, $\mu$ the chemical potential, $v$ the spin-orbit coupling, $\boldsymbol{\sigma}$ the Pauli matrices, $M$ the magnetization, and $\Delta$ the orbital magnetization generated by the BCD from the WTe$_2$ layer. The anomalous Hall conductivity $\sigma_{AHE}$ is obtained by integrating all the states below Fermi level

$$\sigma_{\mathrm{AH}} = \frac{e^2}{h} \int \left[ \frac{v^2(M+\Delta)}{2\left(v^2k^2+(M+\Delta)^2\right)^{\frac{3}{2}}} \right] \left[ n_{\mathrm{F}}(\epsilon_-(k)) - n_{\mathrm{F}}(\epsilon_+(k)) \right] \frac{\mathrm{d}\mathbf{k}}{2\pi} \qquad (3)$$

where $n_{\mathrm{F}}(\epsilon_\pm(k))$ are the Fermi distribution of the lower (−) and upper (+) bands, respectively, and $n_{\mathrm{F}} = \frac{1}{e^{\frac{\epsilon-\mu}{k_{\mathrm{B}}T}}+1}$. We calculate $M$ self-consistently starting from a Hubbard interaction $U$. In principle $\mu$ should also be calculated self-consistently, starting from a given value fixing the electron density at $M = 0$. However, we found the change in $\mu$ to be negligible. At zero temperature, we find that the system undergoes a first order transition at $\Delta = 0$ and $U > U_{\mathrm{cr}}$ where $U_{\mathrm{cr}}$ is the critical Hubbard interaction, to give $M > \mu$, and $\sigma_{\mathrm{AH}}$ is close to the value $e^2/(2h)$. For finite temperatures, as shown in Fig. 3c, a tail is formed by introducing a homogeneous $\Delta$ acting on FGT. This suggests that one can indeed treat the applied charge current as a control parameter for the magnetic phase transition. Note that the $\sigma_{\mathrm{AH}}$ curves in Fig. 3c have discontinuous jumps from metastable states to the energy minima. This is an artefact of the uniform magnetization approximation which does not account for magnetic domain walls and defects in real samples that make the curves continuous. Like in Fig. 3a, one can determine $T_{\mathrm{C}}$ for each $\Delta$ by finding the $T$-axis intercept of the linear fit. As summarised in Fig. 3d, $T_{\mathrm{C}}$ increases monotonically with the magnitude of $|\Delta|$, which is consistent with the experimental results shown in Fig. 3b.

In summary, we have demonstrated a new method to electrically control magnetism by utilizing the current-induced out-of-plane spin accumulation. This approach does not rely on extra charge transfer and is distinctly different from conventional control by using the electric field in capacitors. We show that this approach can boost the Curie temperature of FGT from 200 K to 370 K, and even larger modulation can be expected if a better device sustaining larger current could be fabricated. This method can potentially drive the development of devices made from two-dimensional van der Waals magnets for practical applications. Note that this route is not limited to vdW magnets but can also be applied to other magnetic materials including magnetic metals and (quantum) magnetic insulators. Beyond its effectiveness in enhancing $T_C$, we also show that the electrical current can act as a control parameter to drive magnetic phase transitions. We expect many more exotic current-driven phenomena in other magnetic materials.

**Figure captions**

**Fig. 1 | Schematic of the device, the measurement set up and the *T*-dependent anomalous Hall effect. a**, *bc*-plane-view of the atomic lattice of the WTe$_2$/FGT stack. At the interface, both WTe$_2$ and FGT are terminated by Te atoms, enabling strong hybridization through Te 5p orbitals. **b**, Cross section HRTEM image of the WTe$_2$/FGT heterostructure. **c**, Schematic of the experimental set up for anomalous Hall effect measurements. $I_{ac}$ is the ac sensing current, $I$ is the dc modulation current, and $V_{ac}$ is the detected transverse voltage. **d**, Transverse resistivity $\rho_{xy}$ loop as a function of external magnetic field $H_z$ measured at temperatures between 150 K and 230 K. $\rho_{AH}$ is the anomalous Hall resistivity, which is determined by the loop height. **e**, Temperature dependence of the anomalous Hall resistivity $\rho_{AH}$ for WTe$_2$ (1.6 nm)/FGT(2.4 nm). The inset shows the temperature dependence of $\rho_{AH}$ for pure FGT (7 nm).

**Fig. 2 | Current-induced ferromagnetism at 300 K. a**, Upper panel: time-dependence of the dc modulation current varying from 0 mA to ±0.5 mA. Lower panel: time-dependence of the anomalous Hall conductivity response $\sigma_{AH}$ for WTe$_2$/FGT measured at $T$ = 300 K and $H$ = 0. Here $\sigma_{AH}$ is obtained by $\rho_{AH}/(\rho_{xx}^2+\rho_{AH}^2)$, where $\rho_{xx}$ is the longitudinal resistivity. **b**, *I*-dependence of $\sigma_{AH}$ at $T$ = 300 K. The solid straight line is a guide to the eyes. **c**, The same as (a) but measured for pure FGT. No current-induced

$\sigma_{AH}$ is observed. **d**, **e** and **f**: Schematics of the current-induced magnetism in WTe$_2$/FGT for $I > 0$, $I = 0$ and $I < 0$.

**Fig. 3 | *T*-dependence of anomalous Hall conductivity: experimental results and theoretical simulations. a**, *T*-dependence of $\sigma_{AH}$ measured under $I = \pm 0.5$ mA, $\pm 0.3$ mA, $\pm 0.1$ mA and 0 mA. The dashed lines are the linear fits of the $\sigma_{AH}$-*T* traces for $T > 200$ K, and the Curie temperature $T_C(I)$, denoted by solid points, is determined by the *T*-axis intercept. **b**, *I*-dependence of $T_C$ extracted from (**a**). **c**, Calculated *T*-dependence of $\sigma_{AH}$ using the two-band model for various $\Delta$ values. Here *T* is normalized by $T_C^0$ ($T_C^0$ denotes $T_C$ at $\Delta = 0$). The dashed lines are the linear fits and the Curie temperature $T_C(\Delta)$, denoted by solid points, is determined by the *T*-axis intercept. **d**, $\Delta$-dependence of $T_C$ extracted from (**c**).

**Fig. 4 | Scaling behavior and critical exponents. a**, Temperature dependence of $\sigma_{AH}$ at $I = 0$ and $H = 0$. The solids line is a power-law fit to the data, $\sigma_{AH}(t) \propto (-t)^{\alpha\beta}$. By using $\beta = 0.37$ determined by magnetization measurements, $\alpha = 2.1$ is obtained. **b**, $\sigma_{AH}$ as a function of *I* at $T_C$. The solid line is fitted by $\sigma_{AH}(I) \propto I^{\alpha/\delta}$ and $\delta = 3.7$ is determined. **c**, Scaled AHE conductivity $\sigma_{AH}/I^{\alpha/\delta}$ versus scaled temperature $t/I^{1/\beta\delta}$ for the data sets shown Fig. 3**a**.

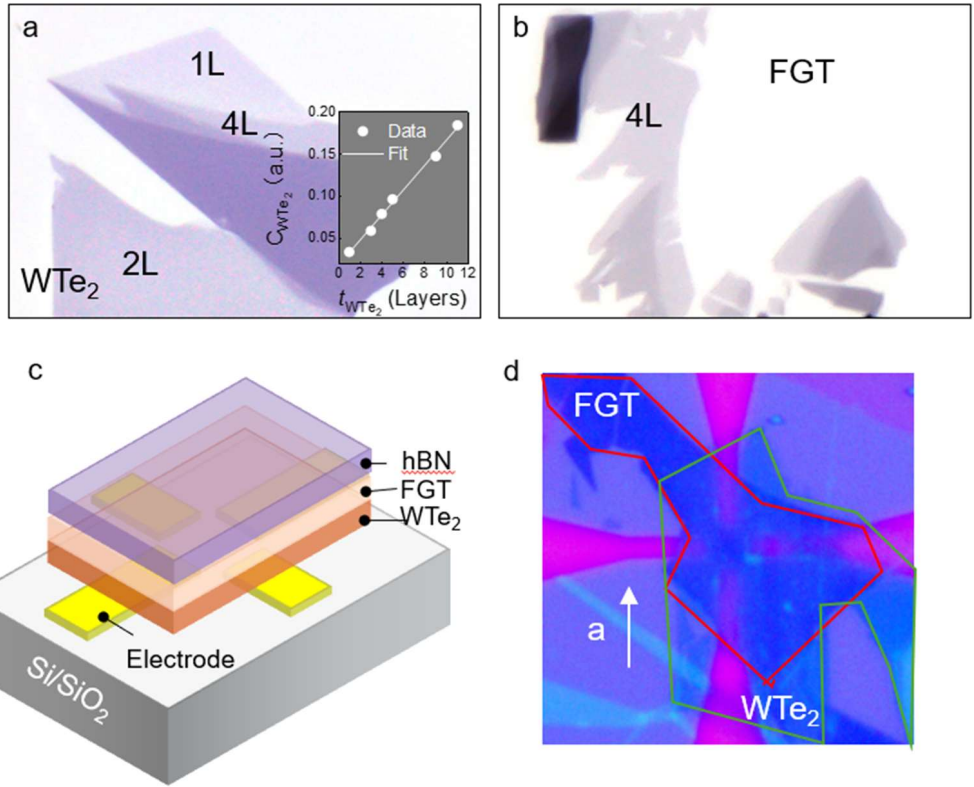

**Extended Data Figure 1 | Device fabrication. a** and **b**: Optical micrograph of WTe$_2$ and FGT flakes. The inset of (**a**) shows the optical contrast of the WTe$_2$ flakes with different layers. **c**, Schematic of the stack. **d**, Optical image of the device.

## Methods

### Materials and device fabrication

Bulk WTe$_2$ and FGT crystals were purchased from HQ Graphene. Extended Figures S2(a) and S2(b) show optical images of a bilayer WTe$_2$ flake and a four-layer FGT flake exfoliated onto a SiO$_2$ (285 nm)/Si substrate. The layer thicknesses were determined from optical contrast mapping [inset of Extended Fig. 1a] and atomic force microscopy, yielding monolayer thicknesses of approximately 0.8 nm for WTe$_2$ and 0.9 nm for FGT. Hexagonal boron nitride (hBN), FGT, and WTe$_2$ flakes were sequentially picked up using a polydimethylsiloxane (PDMS)/polycarbonate (PC) stamp and assembled onto prepatterned cross electrodes [Extended Fig. S1(c)], with the charge current applied along the low-symmetry *a*-axis of WTe$_2$. An optical image of the completed device is shown in Extended Fig. 1d.

**Magneto-transport measurements**

All the electrical measurements were performed in a cryostat equipped with a 3D vector magnet. To measure the AHE hysteresis loop, a low frequency ac current $I_{ac}$ (rms amplitude of 10 μA with a frequency of 31 Hz) is applied between the source and drain contacts by a Keithley 6221 current source, and the transverse voltage $V_{ac}$ is detected by a SR830 lock-in amplifier. The AHE hysteresis loop is measured by sweeping the out-of-plane magnetic field $H_z$, and the transverse resistance $R_{xy}$ is obtained by $R_{xy} = V_{ac}/I_{ac}$. For modulation of $T_C$ measurements, a superposition of $I_{ac}$ and the dc current $I$ is generated using a Keithley 6221 current source (i.e., in wave mode with an offset), and the transverse voltage $V_{ac}$ was measured by a lock-in amplifier; the $R_{xy}$ loop is measured by sweeping $H_z$. While sweeping the magnetic field we keep a constant current $I$ or change its polarity as the magnetization **M** reverses.

**Acknowledgements**

This work was funded by the Deutsche Forschungsgemeinschaft by TRR 360-492547816, by SFB1277-314695032, and by 570332449.

**Author contribution**

L.C. planned the study. J. G., T. M., X. H., Y. G. and J. S. fabricated the devices and collected the data. L.C. and J. G. analysed the data. C. S. and A. O. did the HRTEM measurements. P.R. and J.K. did the calculations as well as the theoretical input. L.C. wrote the manuscript with input from all others. All authors discussed the results.

**Competing financial interests**

The authors declare no competing financial interests.

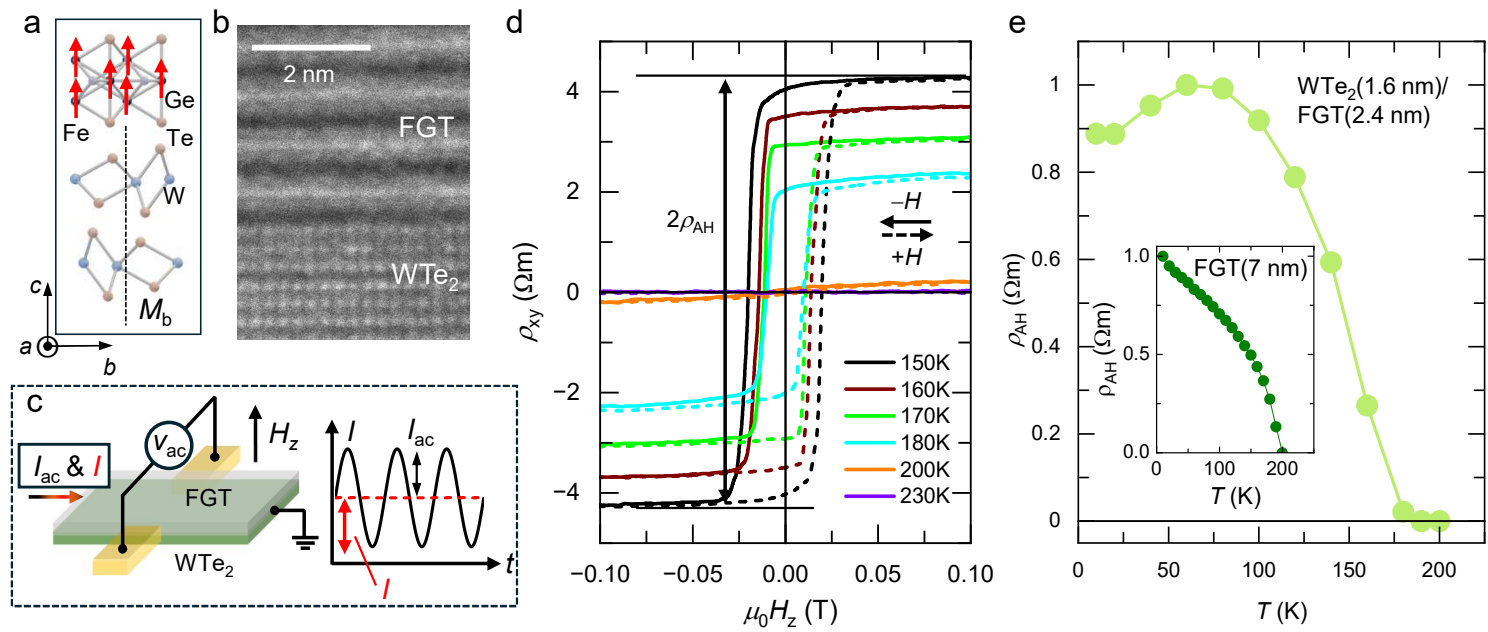

Fig. 1 Guo et al.

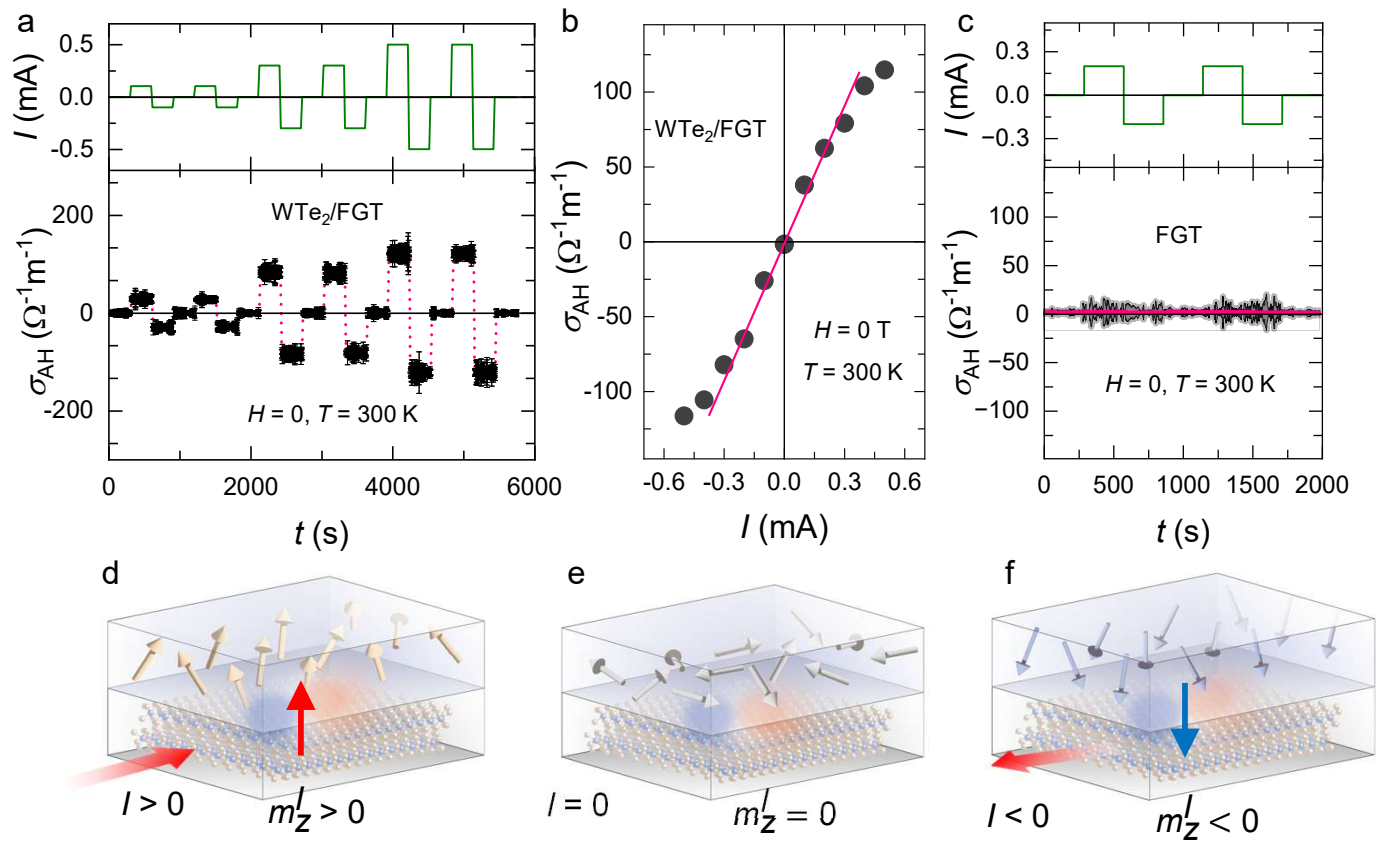

Fig. 2 Guo et al.

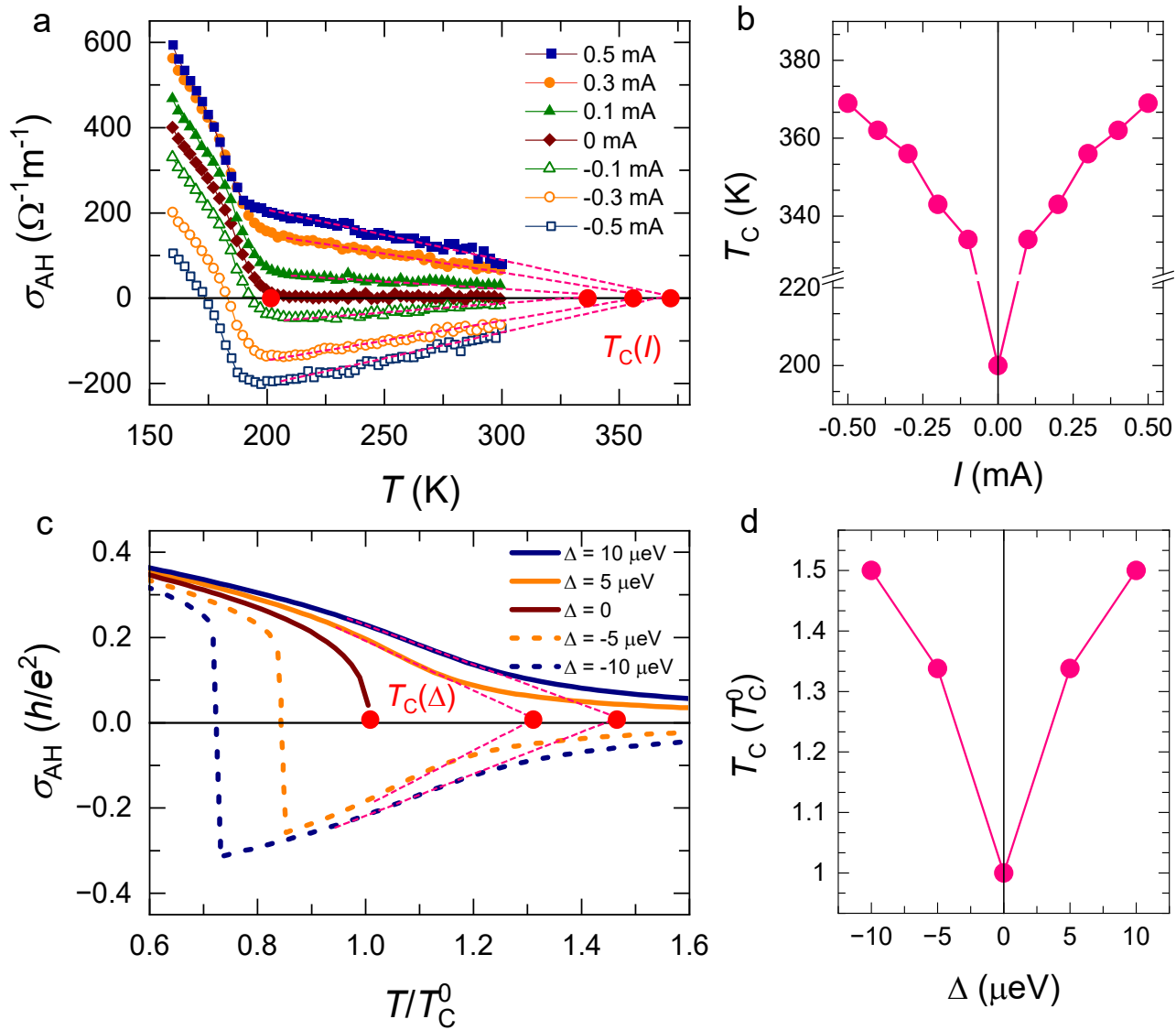

Fig. 3 Guo et al.

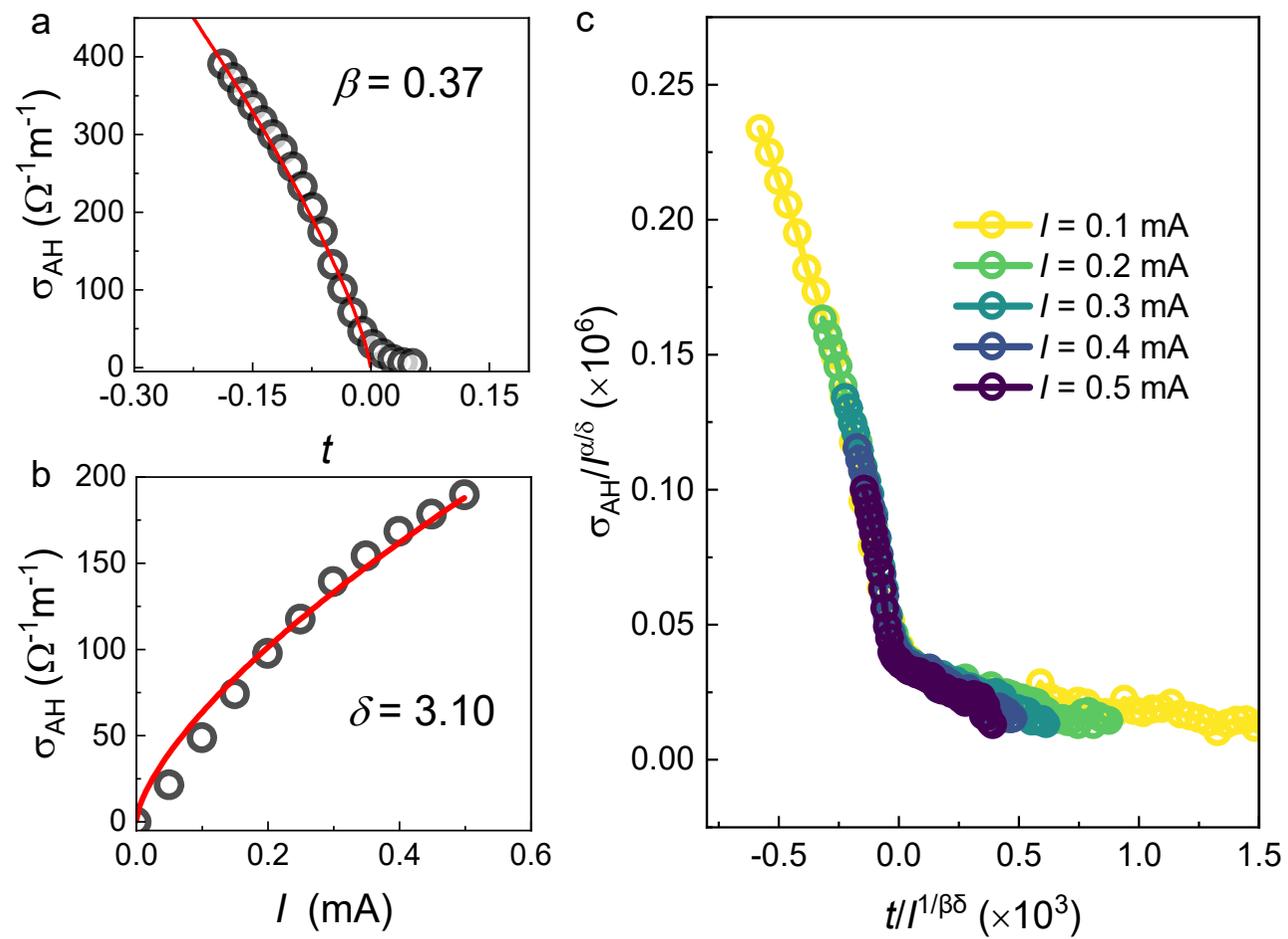

Fig. 4 Guo et al.